\documentclass[fleqn,12pt,twoside]{article}
\usepackage[headings]{espcrc1}
\usepackage{amsmath}

\readRCS
$Id: espcrc1.tex,v 1.2 2004/02/24 11:22:11 spepping Exp $
\usepackage{graphicx}
\usepackage[figuresright]{rotating}

\title{Ultracold atomic quantum gases far from equilibrium}

\author{Thomas Gasenzer\address[UniHD]{Institut f\"ur Theoretische Physik, 
        Ruprecht-Karls Universit\"at Heidelberg,\\
	Philosophenweg 16, 
        69120 Heidelberg, Germany}%
        \thanks{Funded by Deutsche Forschungsgemeinschaft.},
        J\"urgen Berges\address[TUDA]{Institut f\"ur Kernphysik, 
	Technische Universit\"at Darmstadt, \\
	 Schlossgartenstra\ss e 9, 64289 Darmstadt, Germany},
        Michael G. Schmidt\addressmark[UniHD],
        and
        Marcos Seco\addressmark[UniHD]}
       
\runtitle{Ultracold atomic quantum gases far from equilibrium}
\runauthor{T. Gasenzer, J. Berges, M. G. Schmidt, and M. Seco}

\begin{document}

% typeset front matter
\maketitle

%================================================================================
%================================================================================
\begin{abstract}
We calculate the time evolution of a far-from-equilibrium initial state of a non-relativistic ultracold Bose gas in one spatial dimension.
The non-perturbative approximation scheme is based on a systematic expansion of the two-particle irreducible effective action in powers of the inverse number of field components. 
This yields dynamic equations which contain direct scattering, memory and off-shell effects that are not captured in mean-field theory. 
\end{abstract}

%================================================================================
%================================================================================
\section{Non-equilibrium ultracold atomic gases}

In recent years, sophisticated new technologies have been developed for the cooling and trapping of ultracold atomic gases: 
These allow the contact-free storage of atoms in almost arbitrarily shaped "atom traps", as well as the manipulation and Feshbach-resonant enhancement of the interactions between the atoms using externally applied electromagnetic fields  \cite{Stwalley1976b}. %,Tiesinga1992a,Mies2000a}.
(Quasi) one- and two-dimensional traps \cite{Schmiedmayer2000a,Pitaevskii2003a} as well as optical lattices \cite{Jaksch1998a} %,Bloch2004a} 
allow to realize strongly correlated many-body states of atoms reminiscent of similar phenomena in condensed matter systems.
Since these external conditions can be varied very quickly while the experiment is running and the reaction of the gas be monitored extremely precisely, high-precision studies of the quantum dynamics of many-body systems become possible even far from equilibrium. 
%Due to the mentioned freedom in choosing the trap geometry, the atoms can be put into almost arbitrary, but precisely defined, motional patterns. 
%With these technologies, physicists have been given completely new tools with which the investigation of atomic few- and many-body physics is being revolutionized. 
%This adds to the theory of non-equilibrium many-body quantum dynamics being one of the most interesting topics in theoretical physics. 

The 1+1-dimensional systems of identical bosons of mass $m$ to be considered are described by fields obeying $[\hat\Psi(x,t),\hat\Psi^\dagger(y,t)]=\delta(x-y)$. Their dynamics is governed by the Hamiltonian  
\begin{align}
\label{eq:MBHamiltonian}
 &H
 =-\int dx\hat\Psi^\dagger(x)\frac{\hbar^2\partial^2_{x}}{2m}\hat\Psi(x)
 %\nonumber\\
 %&\qquad
 +\ \frac{1}{2}\int dx\,dy\,
     \hat\Psi^\dagger(x)\hat\Psi^\dagger(y)V(|x-y|)
     \hat\Psi(y)\hat\Psi(x).
 %    \\
 %&H_\mathrm{1B}(x)
 %= -\frac{\hbar^2\partial^2_{x}}{2m} + V_\mathrm{trap}(x),
%\label{eq:1BHamiltonian}
\end{align}
At ultralow temperatures, where the thermal de Broglie wave lengths are much larger than the effective range of the binary Born-Oppenheimer potential $V$, this can be approximated by the effective local coupling $V(x)=g_\mathrm{1D}\delta(x)$.
%The field operators obey the bosonic commutation relation $[\hat\Psi(x,t),\hat\Psi^\dagger(y,t)]=\delta(x-y)$ while all other equal-time commutators vanish.

%A strongly interacting gas with $g_\mathrm{1D}=2\hbar^2a/(ml_\perp^2)$ can be obtained in an elongated, e.g., a cylindrical trap with strong transversal harmonic confinement with oscillator length $l_\perp$.
%If the atoms can not pass each other within such a tube, the gas enters the so-called Tonks-Girardeau regime, and the dimensionless interaction parameter $\gamma=g_\mathrm{1D}m/(\hbar^2n)$ is much larger than 1 \cite{Tonks1936a,Girardeau1960a,Paredes2004a}. 

%Conventionally, the dynamics of many-body systems of suffiently weakly interacting particles is described using perturbative approximation schemes for the exact quantum field theoretical many-body equations of motion.
%It is clear that for strongly interacting systems for which the relevant expansion parameter is no longer small, a perturbative approach must eventually fail.

A particular challenge represent systems far from equilibrium for which conventional perturbative and mean-field approximation schemes fail to describe the long-time evolution.
We present a dynamical many body theory of an ultracold Bose gas which systematically extends beyond such approximations.
The non-perturbative approach is based \nolinebreak on

%================================================================================
\begin{figure}[htb]
\begin{minipage}[b]{79mm}
a systematic expansion of the two-particle irreducible (2PI) effective action in powers of the inverse number of field components ${\cal N}$ \cite{Cornwall1974a,Berges2002a,Aarts2002b}. 
This 2PI $1/{\cal N}$ expansion, to next-to-leading order (NLO), yields dynamic equations which allow to describe far-from-equilibrium dynamics as well as the late-time approach to quantum thermal equilibrium.
We emphasize that mean-field approximations fail to describe this dynamics even qualitatively. 
Similarly, standard kinetic descriptions based on two-to-two collisions give a trivial (constant) dynamics because of phase space restrictions in one spatial dimension.
Recently, these methods have allowed important progress in \nolinebreak describ-
\vspace*{-42pt}\\ 
\end{minipage}
%================================================================================
\hfill
%================================================================================
\begin{minipage}[b]{77mm}
\includegraphics[width=77mm]{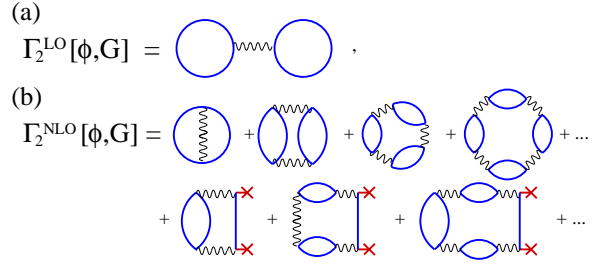}
\vspace*{-13.5mm}\\
\caption{%
Diagrammar of the LO and NLO contributions in the $1/\cal N$-expansion, to the 2PI part $\Gamma_2[\phi,G]$ of the 2PI effective action.
The solid lines represent 2-point functions $G_{ij}(x,y)$, the crosses fields $\phi_i(x)$, and the wiggly lines vertices $g_\mathrm{1D}\delta(x-y)$.
\vspace{2mm}
%At each vertex, it is summed over double field indices $i$ and integrated/summed over double time and space variables $x$.
}
\label{fig:2PINLO1N}
\end{minipage}
%================================================================================
\end{figure}

%================================================================================
%================================================================================
\noindent
\ \vspace*{-1cm}\\
ing the non-equili\-bri\-um dynamics of strongly interacting relativistic systems for bosonic \cite{Berges2002a,Berges2003b,Cooper2003a,Arrizabalaga2004a}, %Mihaila2001a
as well as fermionic degrees of freedom \cite{Berges2003a,Berges2004b}. 
%Our aim is to employ the 2PI effective action for ultracold quantum gases and to numerically solve the 2PI $1/{\cal N}$ expansion to next-to-leading order. 
%This is exemplified for a homogeneous ultracold Bose gas in one spatial dimension. 
%We compute the time evolution of an initial state which is far from thermal equilibrium. 
%After a characteristic short-time scale it is found to be driven to a quasistationary state.
%However, the system is still far from equilibrium and the thermal equilibration  time can exceed the early-time scale by orders of magnitude. 
%In particular, a unique temperature can not be attributed to the intermediate quasistationary state.
The 2PI $1/{\cal N}$ expansion has also been successfully applied to compute critical exponents in thermal equilibrium near the second-order phase transition of a model in the same universality class \cite{Alford2004a}.
For details we refer to \cite{Gasenzer2005a} as well as to \cite{Temme2006a,Berges2006a}.

%================================================================================
\section{The 2PI effective action in NLO 1/${\cal N}$ approximation}

The 2PI effective action~\cite{Luttinger1960a,Baym1962a,Cornwall1974a} is obtained as a double Legendre transform of the generating functional of connected Greens functions and is conveniently written as 
\begin{align}
\label{eq:2PIEAexp}
  \Gamma[\phi,G]
  &= S[\phi] +\frac{i}{2}\,\mathrm{Tr}\left\{\ln G^{-1}+G_0^{-1}[\phi]G\right\} 
     +\Gamma_2[\phi,G]
%     \nonumber\\
%  &\qquad
  +\mathrm{const.}\,,
\end{align} 
which contains the contribution from the classical action $S$, a one-loop-type term and a term $\Gamma_2[\phi,G]$ that contains all the rest.
%The trace, the logarithm and the product of Greens functions in the third term are meant in the functional sense.
$G_0^{-1}$ is the inverse of the classical propagator with
$iG_{0,ij}^{-1}(x,y;\phi)= \delta^2S[\phi]/\delta\phi_i(x)\delta\phi_j(y)$.
We denote the time-space vector by $x=(x_0,x_1)$, etc., where $x_0$ is the time component.
$\phi_i$, $i=1,2$ are the ${\cal N}=2$ components of the macroscopic or mean field, e.g., 
the real and imaginary parts $\phi_1=\sqrt{2}\mathrm{Re}\,\Psi$, $\phi_2=\sqrt{2}\mathrm{Im}\,\Psi$.
Correspondingly, the components of the connected 2-point function are denoted by $G_{ij}(x,y)=\langle{\cal T}\hat\phi_i(x)\hat\phi_j(y)\rangle-\phi_i(x)\phi_j(y)$.
The equations of motion for $\phi_i(x)$ and $G_{ij}(x,y)$ are, finally, derived using the stationarity conditions
%
%\begin{align}
%\label{eq:2PIStatCondPhi}
  ${\delta\Gamma[\phi,G]}/{\delta \phi_i(x)} = 0$,
%  \qquad
%\label{eq:2PIStatCondG}
  ${\delta\Gamma[\phi,G]}/{\delta G_{ij}(x,y)} = 0$.
%\end{align} 
%
$\Gamma_2$ may be represented by the series of all closed 2PI diagrams consisting of full propagators $G$, field insertions $\phi$, and bare vertices proportional to $g_\mathrm{1D}$ \cite{Cornwall1974a}.
%However, in practice, truncations of this series are necessary, as, e.g., a coupling expansion in powers of $g_\mathrm{1D}$ or a loop expansion.
The 2PI $1/{\cal N}$ expansion involves a resummation of the diagrams up to infinite order in $g_\mathrm{1D}$ and reaches substantially beyond the conventional (1PI) $1/\cal N$ approach. 
For the NLO resummation cf.~Fig.~\ref{fig:2PINLO1N}, for explicit expressions of $\Gamma$ and the dynamic equations we refer to \cite{Gasenzer2005a}. 
For the following discussion, it is convenient to express all 2-point correlation functions in terms of their connected statistical and spectral parts,
%
%\begin{align}
%\label{eq:Fijrhoij}
  $F_{ij}(x,y)
  =\langle \{\hat{\Phi}_i(x),\hat{\Phi}_j(y)\}\rangle_c/2$,
%  \qquad
  $\rho_{ij}(x,y)
  =i\langle [\hat{\Phi}_i(x),\hat{\Phi}_j(y)] \rangle_c$,
%\end{align} 
%
such that $G_{ij}(x,y)=F_{ij}(x,y) -({i}/{2})\rho_{ij}(x,y)\,\mathrm{sgn}_{\cal C}(x_0-y_0)$, where $\cal C$ denotes ordering along the closed real-time path.
%Furthermore, $\int_x=\int_{\cal C}dx_0\int x_1$, with the time-integral taken over the Schwinger-Keldysh \cite{Schwinger1961a,Keldysh1964a} contour \cite{Berges2005a}.
%from $x_0=0$ which is taken as the initial time, to the maximum relevant time $t$ and back to $0$.
%This choice of contour can be understood as a way to ensure the normalization of the generating functional.
%\vspace*{-9mm}
%\noindent

%================================================================================
%================================================================================
\section{Equilibration of a homogeneous ultracold 1-dimensional (1D) Bose gas}
\label{sec:results}
We have investigated the dynamic evolution of a 1D Bose gas of sodium atoms in a box of length $L=N_sa_s$ ($N_s=64$ modes on a grid with $a_s=1.33\,\mu$m), with periodic boundary conditions, starting in a non-equilibrium state which is solely described by a Gaussian momentum distribution $n(t=0;p)=N^{-1}\exp\{-p^2/\sigma^2\}$ and normalized such that the line density is fixed to $n_1=10^7$ atoms$/$m.
All higher correlation functions as well as $\phi_i$ are assumed to vanish initially
%, i.e., only the spatial Fourier transforms $F_{ii}(0,0;p)=n(0;p)+1/2$, $i=1,2$, and $-\rho_{12}(0,0;p)=\rho_{21}(0,0;p)=1$, by virtue of the commutation relations, are non-zero 
\cite{Gasenzer2005a}. 
%The numerical grid of $N_s=64$ points is spaced by $a_s=1.33\,\mu$m.
The atoms are weakly interacting with each other, such that $g_\mathrm{1D}=\hbar^2\gamma n_1/m$, with the dimensionless parameter $\gamma=1.5\cdot10^{-3}$. 
%
%=====================================================================================
\begin{figure}[tb]
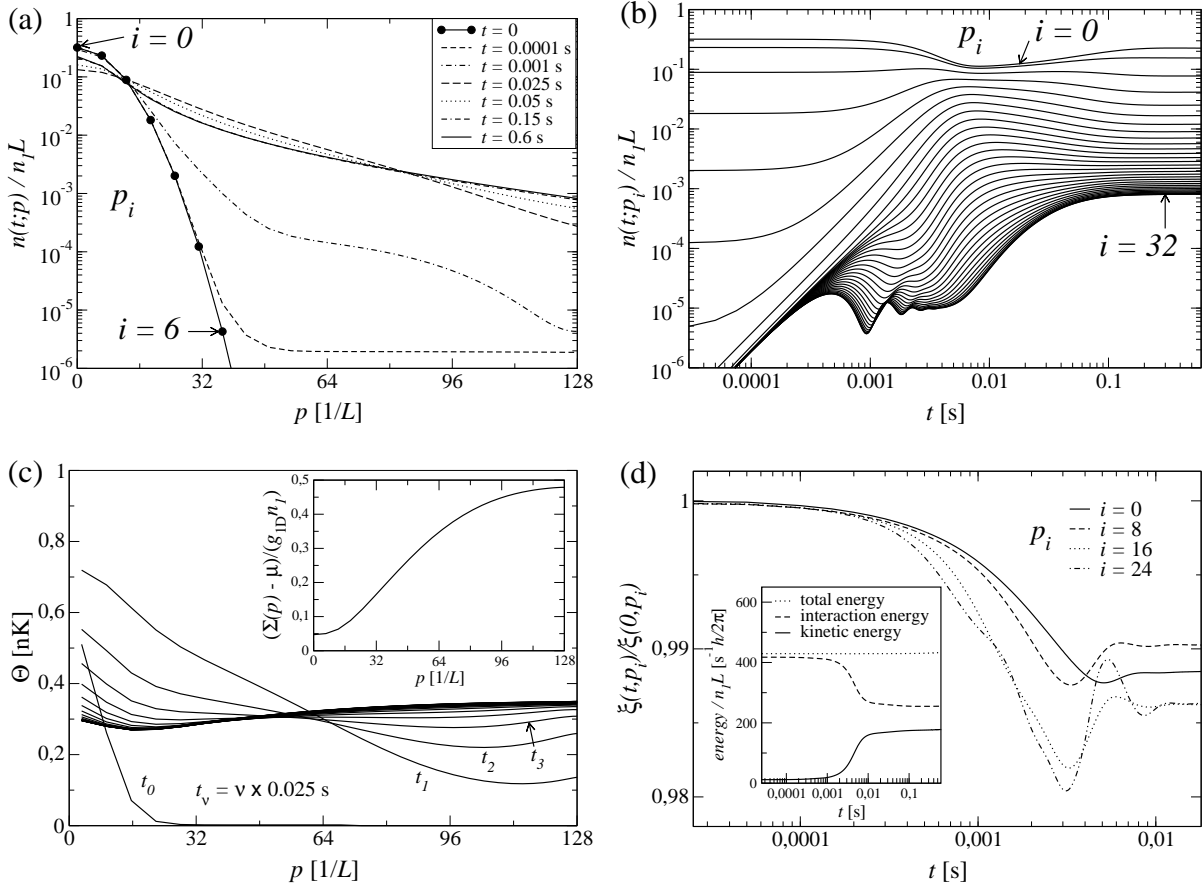

\begin{minipage}[t]{79mm}
\includegraphics[width=78mm]{fig2.eps}
\vspace*{-14mm}\\
%\caption{ }
\label{fig:n1ofp}
\end{minipage}
\hspace{\fill}
\begin{minipage}[t]{79mm}
\includegraphics[width=78mm]{fig3.eps}
\vspace*{-1mm}\\
%\caption{ }
\label{fig:n1oft}
\end{minipage}
%\end{figure}
%\ \vspace*{-10mm}\\
%
%\begin{figure}[htb]
\begin{minipage}[t]{79mm}
\includegraphics[width=78mm]{fig4.eps}
\vspace*{-17mm}\\
%\caption{ }
\label{fig:betaofp}
\end{minipage}
\hspace{\fill}
\begin{minipage}[t]{79mm}
\includegraphics[width=78mm]{fig5.eps}
%\vspace*{-17mm}\\
%\caption{ }
\label{fig:Sigmaofp}
\end{minipage}
\vspace*{-18mm}\\
\caption{Equilibration of ultracold 1D Bose gas. (a) and (b) Time evolution of momentum mode distribution. (c) Development of thermal distribution with temperature $\Theta$. (inset) Momentum dependence of self energy. (d) Establishment of fluctuation-dissipation relation. (inset) Energies. For details confer Section \protect\ref{sec:results}.}
\ \vspace*{-15mm}\\
\label{fig:results}
\end{figure}
%=====================================================================================
%
Figs. \ref{fig:results}a,b show the time evolution of the occupations of the momentum modes $p_i=(2/a_s)\sin(\pi i/N_s)$ as functions of $p$ and $t$, respectively.
Although the system very quickly, after about $5\,\mu$s, evolves to a quasistationary state, the final drift to the equilibrium distribution takes roughly ten times longer.
Note that the mean-field Hartree-Fock (HF) approximation, which only takes into account $\Gamma_2^\mathrm{LO}$ and the $O(g_\mathrm{1D})$-diagram in Fig. \ref{fig:2PINLO1N}b, would conserve exactly all mode occupations and no equilibration would be seen.
We further investigated the different time scales by fitting the distribution to the Bose-Einstein form $n(t;p)=[\exp\{(p^2/2m-\mu)/k_B\Theta(t;p)\}-1]^{-1}$, with a $p$-dependent temperature variable $\Theta(t;p)$.
Fig. \ref{fig:results}c shows $\Theta(t;p)$ for $t=0...0.6\,$s.
Hence, during the quasistationary drift period, no temperature can be attributed to $n(t;p)$, while, for large $t$, $\Theta$ becomes approximately $p$-independent.
%\footnote{%
%To evolve the system to larger times would, given the particular values for $N_s$ and the time step size, require substantially larger computing resources.}
Its variation at small $p$ reflects the modified dispersion relation.
At large $p$, excitations of the gas will be approximately single-particle-like, such that we can associate $\Theta(0.6\,$s$;128/L)$ with the final temperature $T$.
Using this, we deduce the shift $\Sigma(p)$ in the dispersion relation,  $n(0.6\,$s$;p)=[\exp\{(p^2/2m+\Sigma(p)-\mu)/k_BT\}-1]^{-1}$ (inset of Fig. \ref{fig:results}c) and, with the total energy in equilibrium, $E_\mathrm{tot}=\sum_p[p^2/2m+\Sigma(p)/2]n(0.6\,$s$;p)$, which must be equal to the initial one, the chemical potential $\mu=1.08\,g_\mathrm{1D}n_1$.
%We point out that this can not be obtained in HF approximation, which would give a $p$-independent self-energy and shift $\Sigma$. 

We studied the time dependence of the ratio of unequal-time correlation functions, i.e., $\xi(t;p)=[(F_{11}(t,0;p)^2+F_{12}(t,0;p)^2)/(\rho_{11}(t,0;p)^2+\rho_{12}(t,0;p)^2)]^{1/2}/n(t;p)$, Fig. \ref{fig:results}d.
$\xi$ is a measure of the interdependence of the statistical and spectral functions, which, in thermal equilibrium, are connected through the fluctuation-dissipation relation 
%reading, in momentum-frequency representation, 
$F^\mathrm{(eq)}(\omega,p)=-i[n(\omega,T)+1/2]\rho^\mathrm{(eq)}(\omega,p)$.
One finds that already during the quasistationary drift period, $\xi$ becomes approximately time-independent for all momenta $p_i$, i.e., there is a fluctuation-dissipation relation even though the system is still away from equilibrium.
A similar signature is found when comparing the kinetic and interaction contributions to the total energy as shown in the inset of Fig.~\ref{fig:results}d.
During the drift period, these contributions are constant and approximately equal to each other, calling in mind the virial theorem.
In summary, during the drift period, the system is not yet in equilibrium as far as the momentum distribution and temperature is concerned, but shows important characteristics of a system close to equilibrium.
The details of this is a subject of future work.

\bibliographystyle{report_notitle}
\bibliography{bibtex/mybib,bibtex/additions}

%\begin{thebibliography}{9}
%\bibitem{Scho70} S. Scholes, Discuss. Faraday Soc. No. 50 (1970) 222.
%\bibitem{Mazu84} O.V. Mazurin and E.A. Porai-Koshits (eds.),
%                 Phase Separation in Glass, North-Holland, Amsterdam, 1984.
%\bibitem{Dimi75} Y. Dimitriev and E. Kashchieva, 
%                 J. Mater. Sci. 10 (1975) 1419.
%\bibitem{Eato75} D.L. Eaton, Porous Glass Support Material,
%                 US Patent No. 3 904 422 (1975).
%\end{thebibliography}

\end{document}